\documentstyle[12pt]{article}

\newcommand{\eq}{\begin{equation}}
\newcommand{\eqa}{\begin{eqnarray}}
\newcommand{\en}{\end{equation}}
\newcommand{\ena}{\end{eqnarray}}
\newcommand{\enn}{\nonumber \end{equation}}

\def\square{\,\lower0.9pt\vbox{\hrule \hbox{\vrule height 0.2 cm
\hskip 0.2 cm \vrule height 0.2 cm}\hrule}\,}

\def\sk{\vskip .4cm}
\def\noi{\noindent}
\def\om{\omega}
\def\al{\alpha}
\def\be{\beta}

\def\Cb{\bar{C}}

\def\epsi{\varepsilon}

\def\we{\wedge}

\def\de{\delta}

\def\part{\partial}

\def\R#1#2{ R^{#1}_{~~~#2} }

\def\Rh{{\hat R}}

\def\Rhat#1#2{ \Rh^{#1}_{~~~#2} }
\def\L#1#2{ \La^{#1}_{~~~#2} }

\def\La{\Lambda}

\def\Cb{{\bf C}}

\def\C#1#2{ {\bf C}_{#1}^{~~~#2} }
\def\c#1#2{ C_{#1}^{~~~#2} }

\def\f#1#2{ f^{#1}_{~~#2} }

\def\M#1#2{ M_{#1}^{~#2} }
\def\qm{q^{-1}}

\def\D{\Delta}

\def\Dp{\Delta^{\prime}}
\def\Ip{I^{\prime}}
\def\ep{\epsi^{\prime}}

\def\qone{q \rightarrow 1}

\def\qh{q^{1\over 2}}
\def\qmh{q^{-{1\over 2}}}

\def\qm{q^{-1}}
\def\qmt{q^{-2}}

\def\n2{{{N+1} \over 2}}

\def\lie#1{\ell_{t_{#1}}}
\def\con#1{i_{t_{#1}}}

\def\Lthree#1#2#3#4{\Lambda^{#1~#2}_{~~~~#3~#4}}
\def\Cthree#1#2#3{{\bf C}_{#1~#2}^{~~~~~#3}}

\def\RB#1#2{{\bf R}^{#1}_{~#2}}

\begin{document}

\begin{titlepage}
\rightline{DFTT-01/94}
\vskip 2em
\begin{center}{\bf  THE LAGRANGIAN OF $q$-POINCAR\'E GRAVITY}
\\[6em]
 Leonardo Castellani
\\[2em]
{\sl Istituto Nazionale di
Fisica Nucleare, Sezione di Torino
\\and\\Dipartimento di Fisica Teorica\\
Via P. Giuria 1, 10125 Torino, Italy.}  \\
\sk

\vskip 2cm

\end{center}
\begin{abstract}
The gauging of the $q$-Poincar\'e algebra
of ref. \cite{qpoincarebic}
yields a non-commutative generalization of
the Einstein-Cartan lagrangian.

We prove its invariance under local $q$-Lorentz rotations and,
up to a total derivative, under $q$-diffeomorphisms. The
variations of the fields are given by their
$q$-Lie derivative, in analogy with the $q=1$ case. The
algebra of $q$-Lie derivatives is shown to close with
field dependent structure functions.

The equations of motion are found, generalizing the
Einstein equations and the zero-torsion condition.

\end{abstract}

\vskip 4.5cm

\noi DFTT-01/94

\noi hep-th/9402033

\noi February 1994
\vskip .2cm
\noi \hrule
\vskip.2cm
\hbox{\vbox{\hbox{{\small{\it e-mail addresses:}}}\hbox{}}
 \vbox{\hbox{{\small decnet=31890::castellani;}}
\hbox{{\small internet= castellani@to.infn.it }}}}

\end{titlepage}

\newpage
\setcounter{page}{1}

%\sect{Introduction }
We describe in this Letter a geometric procedure to gauge
the quantum Poincar\'e algebra found in ref. \cite{qpoincarebic}.
The lagrangian we obtain is a generalization of the
Einstein-Cartan lagrangian, and has the same kind of symmetries
(now $q$-deformed symmetries) as its classical counterpart:
it is invariant under local Lorentz rotations and $q$-diffeomorphisms.
\sk
As one could expect, the differential calculus on
the $q$-deformed Poincar\'e group is the correct
framework for the program of finding a $q$-generalization
of Einstein gravity. It was not obvious that this
program could be carried to the end: in fact it can be done.
We refer to \cite{qpoincarebic} for most of the technicalities
regarding the  inhomogeneous quantum groups $ISO_q(N)$
and their differential calculus. Here we concentrate
directly on the $ISO_q(3,1)$ quantum Lie algebra, and
discuss its gauging. The method we follow is a natural
$q$-extension of the geometric procedure described in
\cite{Cas2} for classical gauge and (super)gravity theories \footnote
{the so-called ``group manifold approach" was initiated in ref.s
\cite{Regge}.}.
\sk
The starting point is the
$q$-algebra $ISO_q(3,1)$ of ref. \cite{qpoincarebic}:
\eqa
& &[\chi_{ab},\chi_{cd}]=C_{bc} \chi_{ad} + C_{ad} \chi_{bc}-
         C_{bd} \chi_{ac}-C_{ac} \chi_{bd} \label{lorentz}\\
& &{}\nonumber\\
& &[\chi_{12}, \chi_a]_{q^{-1}} =\qmh C_{2a} \chi_1-\qmh C_{1a} \chi_2
       \nonumber \\
& &[\chi_{13}, \chi_a]_{q^{-1}} =\qmh C_{3a} \chi_1-\qmh C_{1a} \chi_3
       \nonumber \\
& &[\chi_{14}, \chi_a] =C_{4a} \chi_1-C_{1a} \chi_4
       \nonumber \\
& &[\chi_{23}, \chi_a] =C_{3a} \chi_2-C_{2a} \chi_3
       \nonumber \\
& &[\chi_{24}, \chi_a]_q =\qh C_{4a} \chi_2-\qh C_{2a} \chi_4
       \nonumber \\
& &[\chi_{34}, \chi_a]_q =\qh C_{4a} \chi_3-\qh C_{3a} \chi_4
       \label{mixed}\\
& &{}\nonumber\\
& &[\chi_1,\chi_2]_{\qm}=0,~~~~~~~[\chi_1,\chi_3]_{\qm}=0 \nonumber\\
& &[\chi_1,\chi_4]_{\qmt}=0,~~~~~~~[\chi_2,\chi_3]=0 \nonumber\\
& &[\chi_2,\chi_4]_{\qm}=0,~~~~~~~[\chi_3,\chi_4]_{\qm}=0 \label{momenta}
\ena
where $[A,B]_s \equiv AB-sBA$. The subalgebra spanned by the Lorentz
generators $\chi_{ab}$ ($= - \chi_{ba}$) is {\sl classical};
 the deformation parameter
$q$ appears only in  the commutation relations (\ref{mixed}) and
 (\ref{momenta}), involving the momenta $\chi_a$.
The metric $C_{ab}$ is given by
\eq
C_{ab}=\left(  \begin{array}{cccc}
     0&0&0&1\\
     0&1&0&0\\
     0&0&1&0\\
     1&0&0&0\\
      \end{array} \right)  \label{lorentzmetric}
\en
\noi with  Lorentz signature $(+,+,+,-)$. Only
in the classical limit $\qone$
can one redefine the generators so as to
diagonalize (\ref{lorentzmetric}).
The fact that
the metric is diagonal in the
indices 2,3  (and not completely antidiagonal
as for the $q$-groups defined in \cite{FRT}) is
due to the existence of a
particular real form on $SO_q(4;{\bf C})$. This real form, first
discussed in ref. \cite{Firenze1} for the uniparametric  $q$-groups
$SO_q(2n;{\bf C})$,  was extended to the multiparametric case
and to $ISO_q(2n,{\bf C})$ in \cite{qpoincarebic}, and
allows to redefine antihermitian (linear combinations of the) generators,
bringing  the
antidiagonal metric of ref. \cite{FRT}
in the hybrid form (\ref{lorentzmetric}).
\sk
The algebra (\ref{lorentz})-(\ref{momenta})
was obtained in ref. \cite{qpoincarebic}
via a consistent projection
from the $q$-Lie algebra of a particular
multiparametric deformation of $SO(6)$, for
which the $R$ matrix takes a very simple form: it is diagonal
and satisfies $\Rh^2=1$, with $\Rh \equiv PR$
($\Rhat{ab}{cd} \equiv \R{ba}{cd}$).
\sk
The  $q$-Lie algebra (\ref{lorentz})-(\ref{momenta}) has the form
\eq
\chi_i \chi_j-\L{kl}{ij} \chi_k \chi_l = \C{ij}{k} \chi_k \label{qLie}
\en
\noi where i,j...are adjoint indices
running on the 10 values corresponding
to the indices (a,ab) of the generators of
$ISO_q(3,1)$. The non-vanishing
components of the braiding
matrix $\La$  and
the structure constants $\Cb$, implicitly defined by
(\ref{lorentz})-(\ref{momenta}), are given
below (no sum on repeated indices):
\eqa
& &\Lthree{ab}{cd}{ef}{gh}=\de^a_g \de^b_h
\de^c_e \de^d_f  \label{lthree}\\
& &\Lthree{a}{bc}{de}{f}=(\al_{de})^2 ~\de^a_f
\de^b_d \de^c_e \label{ltwo}\\
& &\Lthree{bc}{a}{f}{de}=(\al_{de})^{-2} ~\de^a_f
\de^b_d \de^c_e \label{ltwobis}\\
& &\L{ab}{cd}=\be_{cd} ~\de^a_d \de^b_c \label{lone}
\ena
\eqa
& &\Cthree{ab}{cd}{ef}={1\over 4} [C_{ad}\de^e_b \de^f_c+
C_{bc} \de^e_a \de^f_d-C_{ac}\de^e_b \de^f_d-
C_{bd} \de^e_a \de^f_c]-(e \leftrightarrow f) \label{cthree}\\
& &\Cthree{ab}{c}{d}={1 \over 2} \al_{ab}~(C_{bc}
\de^d_a-C_{ac} \de^d_b) \label{ctwo}\\
& &\Cthree{c}{ab}{d}=-{1 \over 2}\al_{ab}^{-1}
 ~(C_{bc} \de^d_a-C_{ac} \de^d_b)
\label{ctwobis}
\ena
\noi with
\eqa
& &\al_{12}=\al_{13}=q^{-{1\over 2}},~~\al_{24}=\al_{34}=q^{1\over 2},~~
\al_{14}=\al_{23}=1 \label{alpha}
\ena
\eqa
& &\be_{12}=\be_{13}=\be_{24}=\be_{34}=q^{-1},~~
\be_{14}=q^{-2},~\be_{23}=1,~~\be_{ab}=\be_{ba}^{-1}
\ena
\noi Note that the $\Lambda$ tensor has unit square, i.e.
\eq
\L{ij}{kl} \L{kl}{mn}=\de^i_m \de^j_n
\en
\noi so that the algebra in (\ref{lorentz})-(\ref{momenta}) is
a {\sl minimal} deformation of $ISO(3,1)$. Deformations of
Lie algebras whose
braiding matrix has unit square
were considered some time
ago by Gurevich \cite{Gurevich}.

The $\Lambda$ and $\Cb$ components in (\ref{lthree})-(\ref{ctwobis})
satisfy the following conditions \footnote{this
can be checked directly without too much effort. In \cite{qpoincarebic}
a general proof is given for all $ISO_q(N)$ obtained by the
projective method.}
\eqa
& & \C{ri}{n} \C{nj}{s}-\L{kl}{ij} \C{rk}{n} \C{nl}{s} =
\C{ij}{k} \C{rk}{s}
{}~~\mbox{({\sl q}-Jacobi identities)} \label{bic1}\\
& & \L{nm}{ij} \L{ik}{rp} \L{js}{kq}=\L{nk}{ri} \L{ms}{kj}
\L{ij}{pq}~~~~~~~~~\mbox{(Yang--Baxter)} \label{bic2}\\
& & \C{mn}{i} \L{ml}{rj} \L{ns}{lk} + \L{il}{rj} \C{lk}{s} =
\L{pq}{jk} \L{is}{lq} \C{rp}{l} + \C{jk}{m} \L{is}{rm}
\label{bic3}\\
& & \C{rk}{m} \L{ns}{ml} = \L{ij}{kl} \L{nm}{ri} \C{mj}{s}
\label{bic4}
\ena
These are the ``bicovariance conditions", see ref.s
\cite{Wor,Bernard,Aschieri1},
necessary for the existence of a  bicovariant
differential  calculus (see also the discussion in Appendix
B of \cite{qpoincarebic}).   Whenever we have a set
of matrices $\L{ij}{kl}$, $\C{ij}{k}$ satisfying
(\ref{bic1})-(\ref{bic4}) we can construct a differential
calculus on the quantum group $Fun_q(\M{i}{j})$,
generated by the elements (adjoint representation of the $q$-groups)
$\M{i}{j}$ satisfying
the ``$\Lambda M M$" relations:
\eq
\M{i}{j} \M{r}{q} \L{ir}{pk}=\L{jq}{ri} \M{p}{r} \M{k}{i} \label{LMM}
\en
Consistency of these relations is ensured
by the  QYB equations (\ref{bic2}).
 One can define in the usual way a coproduct $\D(\M{i}{j})=
\M{i}{k} \otimes \M{k}{j}$ and a counit $\epsi (\M{i}{j})=\de^j_i$.
When $\Lambda^2=1$ one can also define a coinverse $\kappa (\M{i}{j})$
with $\kappa^2=1$ (This is done by enlarging the
algebra $Fun_q(\M{i}{j})$,
see Appendix B of \cite{qpoincarebic}).

The generators $\chi_i$ of the $q$-Lie algebra
(\ref{qLie}) are functionals on $Fun_q(\M{i}{j})$:
\eq
\chi_j (\M{i}{k})=\C{ij}{k}
\en
We recall that
products of functionals are defined via the coproduct $\D$, i.e.
$\chi_i \chi_j \equiv (\chi_i \otimes \chi_j) \D$, whereas
functionals act on products as $\chi_i (ab)=
\Dp (\chi_i) (a \otimes b)$, $a,b \in Fun_q(\M{i}{j})$ (see below
 the definition of $\Dp$).

Next we introduce new
 functionals $\f{i}{l}$ via their action on the basis $\M{k}{j}$:
\eq
\f{i}{l} (\M{k}{j})=\L{ij}{kl}
\en
\noi The co-structures of $\chi$ and $f$ are given by:
\eqa
& & \Dp (\chi_i)=\chi_j \otimes \f{j}{i} + \Ip \otimes \chi_i \\
& & \ep (\chi_i)=0 \\
& & \kappa^{\prime}
(\chi_i)= - \chi_j \kappa^{\prime} (\f{j}{i}) \label{chiantipode}
\ena
\eqa
& & \Dp (\f{i}{j})=\f{i}{k} \otimes \f{k}{j} \\
& & \ep(\f{i}{j})=\de^i_j \\
& & \kappa^{\prime} (\f{i}{j})= \f{i}{j} \circ \kappa
\ena

The algebra generated by the $\chi$ and $f$ is a Hopf algebra (the
$\chi , f$ and $f,f$ commutations are given in
 \cite{Bernard,Aschieri1,qpoincarebic}),
 and defines
a bicovariant differential calculus on the $q$-group generated by
the $\M{i}{j}$ elements. For example, one can introduce left-invariant
one-forms $\om^i$ as duals to the  ``tangent vectors" $\chi_i$,
an exterior product
\eq
 \om^i \we \om^j \equiv \om^i \otimes \om^j -
\L{ij}{kl} \om^k \otimes \om^l ,\label{wedge}
\en
\noi an exterior derivative
on $Fun_q(\M{i}{j})$ as
\eq
da= (id \otimes \chi_i) \D (a) \om^i, \mbox{~~~$a \in Fun_q(\M{i}{j})$  }
\en
\noi and so on.  The commutations between one-forms and elements
 $a \in Fun_q(\M{i}{j})$ are given by:
\eq
\om^i a=(id \otimes \f{i}{j}) \D (a) \label{omb}
\en
The exterior derivative can be extended to the (left-invariant)
one-forms via the
deformed Cartan-Maurer equations \cite{Wor,Aschieri1}
\eq
d \om^i + \c{jk}{i} \om^j \we \om^k=0
\en
\noi The $C$ structure constants
appearing in the Cartan-Maurer equations are related
to the $\Cb$ constants of the $q$-Lie algebra as \cite{Aschieri1}:
\eq
\C{jk}{i}=\c{jk}{i}-\L{rs}{jk} \c{rs}{i}
\en
\noi In the particular case $\Lambda^2=I$ it is not difficult to
see that
$C= {1 \over 2} \Cb$.
\sk
The procedure we have advocated in ref.s \cite{Cas1} for the
``gauging" of quantum groups essentially retraces the steps
of the group-geometric method for the
gauging of usual Lie groups, described
for instance in ref.s \cite{Cas2}.

We consider one-forms $\om^i$ which are not left-invariant any
more, so that the Cartan-Maurer equations are replaced by:
\eq
R^i=d \om^i + \c{jk}{i} \om^j \we \om^k \label{curvature}
\en
\noi where the curvatures $R^i$ are now non-vanishing, and satisfy
the $q$-Bianchi identities:
\eq
dR^i-\c{jk}{i} R^j \we \om^k + \c{jk}{i} \om^j \we R^k=0 \label{bianchi}
\en
\noi due to the Jacobi identities on the structure constants
 $\c{ij}{k}$
\cite{Aschieri1}. As in the classical case we can write the
 $q$-Bianchi identities as $\nabla R^i=0$ (these define the
covariant derivative $\nabla$).

Consider now the definition (\ref{curvature})
of the curvature $R^i$, and apply
it to the $q$-Poincar\'e algebra of  (\ref{lorentz})-(\ref{momenta}): the
one-forms dual to $\chi_{ab}$, $\chi_{a}$ are respectively denoted by
$\om^{ab},V^a$ and the corresponding
curvatures read (we omit wedge symbols):
\eq
R^{ab}=d \om^{ab} +  C_{cd}\om^{ac}  \om^{db} \label{Lorentzcurv}
\en
\eq
R^a=dV^a + \al_{af} C_{fb} \om^{af} V^b \label{torsion}
\en
\noi where  $V_a \equiv  C_{ab} V^b$, $\al_{af}$ and $C_{ab}$
are given in (\ref{alpha}) and (\ref{lorentzmetric}), and we used
$\c{ij}{k}={1\over 2} \C{ij}{k}$. We have also rescaled
$\om^{ab}$ by a factor ${1\over 2}$ to obtain standard
normalizations.  $R^{ab}$ is the
$q$-Lorentz curvature, coinciding with the classical
one (as a function of
$\om^{ab}$), and $R^a$ is the $q$-deformed torsion.
\sk
The Bianchi identities, deduced from (\ref{bianchi}), are:
\eq
dR^{ab} - C_{fe} R^{af} \om^{eb} + C_{fe} \om^{af} R^{eb}=0
 \label{Rabbianchi}
\en
\eq
dR^a+ \al_{af} C_{fb} R^{af} V^b-\al_{af} C_{fb} \om^{af} R^b=0
\en
{}From the definition (\ref{wedge}) of the exterior product we see that
for $\Lambda^2=I$ the one-forms $\om^i$  $q$-commute as:
\eq
\om^i \om^j =-\L{ij}{kl} \om^k \om^l
\en
Inserting the $\Lambda$ tensor corresponding
to  (\ref{lthree})-(\ref{lone})
we find:
\eqa
& &V^a \om^{12}=-\qm \om^{12} V^a \nonumber\\
& &V^a \om^{13}=-\qm \om^{13} V^a \nonumber\\
& &V^a \om^{14}=-\om^{14} V^a \nonumber\\
& &V^a \om^{23}=-\om^{23} V^a \nonumber\\
& &V^a \om^{24}=-q \om^{24} V^a\nonumber\\
& &V^a \om^{34}=-q \om^{34} V^a \label{Vomcomm}
\ena
\eqa
& &V^2 V^1 = - \qm V^1 V^2 \nonumber\\
& &V^3 V^1 = - \qm V^1 V^3 \nonumber\\
& &V^4 V^1 = - \qmt V^1 V^4 \nonumber\\
& &V^3 V^2 = - V^2 V^3 \nonumber\\
& &V^4 V^2 = - \qm V^2 V^4 \nonumber\\
& &V^4 V^3 = - \qm V^3 V^4
\ena
\noi and usual anticommutations between the $\om^{ab}$ (components
of the Lorentz spin connection). The exterior product of two identical
one-forms vanishes (this is not
true in general when $\Lambda^2 \not= I$). As a consequence, the
exterior product of five vielbeins is zero.
\sk
We are now ready to write the lagrangian for the
$q$-gravity theory based on $ISO_q(3,1)$. The
lagrangian looks identical to the classical one, i.e.:
\eq
{\cal L}=R^{ab} V^c V^d \epsi_{abcd} \label{lagrangian}
\en
The Lorentz curvature $R^{ab}$, although defined as in the classical
case, has non-trivial commutations with the $q$-vielbein:
\eqa
& &V^a R^{12}=\qm R^{12} V^a \nonumber\\
& &V^a R^{13}=\qm R^{13} V^a \nonumber\\
& &V^a R^{14}=R^{14} V^a \nonumber\\
& &V^a R^{23}= R^{23} V^a \nonumber\\
& &V^a R^{24}= q R^{24} V^a \nonumber\\
& &V^a R^{34}=q  R^{34} V^a
\ena
\noi deducible from the definition
(\ref{Lorentzcurv}). As in ref. \cite{Cas1,Aschieri1},
we make the assumption that the commutations of $d\om^i$ with
the one-forms $\om^l$ are the same as those of $\c{jk}{i} \om^j \om^k$
with $\om^l$, i.e. the same as those valid for $R^i=0$.
For the definition of $\epsi_{abcd}$ in (\ref{lagrangian})
see below.
\sk
We discuss now the notion of $q$-diffeomorphisms. It is known
that there is a consistent $q$-generalization of the Lie derivative
(see ref.s \cite{Aschieri1,Aschieri2,Schupp,qpoincarebic} )
which can be
expressed as in the classical
case as:
\eq
\ell_V = i_V d + d i_V \label{liederivative}
\en
\noi where $i_V$ is the $q$-contraction operator defined
in ref.s \cite{Aschieri1,Aschieri2}, with the following properties:
\eqa
& &i)~~~i_V (a)=0, \mbox{~~$a \in A$,
$V$ generic tangent vector} \nonumber\\
& &ii)~~\con{i} \om^j=\de^j_i I \nonumber\\
& &iii)~\con{i} (\theta \we \om^k)=
\con{r} (\theta) \om^l \L{rk}{li}+(-1)^p~ \theta ~\de^k_i,
\mbox{~~$\theta$ generic $p$-form}
\nonumber\\
& &iv)~~i_V(a \theta + \theta')=a i_V (\theta)+i_V \theta',
\mbox{~~$\theta, \theta' $generic forms}\nonumber\\
& &v)~~~i_{\lambda V} = \lambda i_V,
\mbox{~~$\lambda \in \Cb$}\nonumber\\
& &vi)~~i_{\epsi V} (\theta)=i_V
(\theta) \epsi, \mbox{~~$\epsi \in A$}
\label{conprop}
\ena
As a consequence, the $q$-Lie derivative satisfies:
\eqa
& &i)~~~\ell_V a = i_V (da) \equiv V(a)
\nonumber\\
& &ii)~~\ell_V d =d \ell_V  \nonumber\\
& &iii)~\ell_V (\lambda \theta +
\theta')=\lambda \ell_V (\theta)+ \ell_V (\theta')
\nonumber\\
& &iv)~~\ell_{\epsi V} (\theta)=(\ell_V
\theta)\epsi - (-1)^p  i_V(\theta) d\epsi,
\mbox{~~$\theta$ generic p-form} \nonumber\\
& &v)~~~\ell_{ t_i}(\theta \we \om^k)=
(\lie{r} \theta) \we \om^l \L{rk}{li} +
\theta \we \lie{i} \om^k   \label{lieprop}
\ena
In analogy with the classical case, we define
the $q$-diffeomorphism variation of the fundamental field $\om^i$
as
\eq
\delta \om^k \equiv \ell_{\epsi^i t_i}  \om^k
\en
\noi where according to iv) in (\ref{lieprop}):
\eq
\ell_{\epsi^i t_i} \om^k= (\con{i}
d \om^k+d \con{i} \om^k) \epsi^i + d\epsi^k=
(\con{i} d\om^k )\epsi^i+ d \epsi^k
\en
As in the classical case, there is a
suggestive way to write this variation:
\eq
\ell_{\epsi^i t_i} \om^k= i_{\epsi^i t_i} R^k+ \nabla \epsi^k
\label{varom}
\en
\noi where
\eqa
& &\nabla \epsi^k \equiv d\epsi^k -
\c{rs}{k} \con{i} (\om^r \we \om^s) \epsi^i=
\nonumber\\
& &~~~~~~d\epsi^k-\c{rs}{k} \epsi^r \om^s+\c{rs}{k} \om^r \epsi^s
\ena
\noi Proof: use the Bianchi
identities (\ref{bianchi}) and iii) in (\ref{conprop}).
\sk
\noi Notice that if we postulate:
\eqa
& &\L{rk}{li} \om^l \epsi^i = \epsi^r \om^k \nonumber\\
& &\L{rk}{li} \om^l \we d\epsi^i = -d\epsi^r \we \om^k \label{epsiom}
\ena
\noi we find
\eq
\de (\om^j \we \om^k)=\de\om^j \we
\om^k + \om^j  \we \de \om^k \label{omvar}
\en
\noi i.e. a rule that any ``sensible"
variation law should satisfy. To prove
(\ref{omvar}) use iv) and v) of (\ref{lieprop}). The $q$-commutations
(\ref{epsiom}) were already proposed in \cite{Cas1} in the context
of $q$-gauge theories. A consequence of (\ref{epsiom}) are the
following commutations between the variation parameter and
$d\om^i$:
\eq
\L{rk}{li} d\om^l \epsi^i = \epsi^r d \om^k \label{epsidom}
\en
As discussed in ref.s \cite{Cas1}, it is
consistent to postulate that $R^i$ has the same
commutations with $\epsi^j$ as $d\om^i$:
\eq
\L{rk}{li} R^l \epsi^i = \epsi^r  R^k \label{epsiR}
\en
\sk
We have now all the tools we need to investigate the invariances of the
$q$-gravity lagrangian (\ref{lagrangian}).
The result will be analogous to the classical one:
after imposing the horizontality conditions
\eq
\con{ab} R^{cd}=\con{ab} R^c=0 \label{horizontality}
\en
along the Lorentz directions one finds that, {\sl provided}
the $\epsi$ tensor in  (\ref{lagrangian}) is appropriately defined,
the lagrangian is invariant under $q$-diffeomorphisms
and local Lorentz rotations. Note that, as in the $q=1$ case,
the horizontality conditions (\ref{horizontality})
can be obtained as field equations (see later).
\sk
We first consider Lorentz rotations. Under
these, the curvature $R^{ab}$ and
the vielbein $V^c$ transform as:
\eq
\de R^{ab}\equiv \ell_{\epsi^{gh} t_{gh}}
R^{ab}=\Cthree{ef}{gh}{ab} R^{ef} \epsi^{gh}-
\Cthree{gh}{ef}{ab} \epsi^{gh} R^{ef}\label{varR}
\en
\eq
\de V^c\equiv  \ell_{\epsi^{gh} t_{gh}} V^c=
 -\Cthree{ef}{g}{c} \epsi^{ef} V^g+
\Cthree{g}{ef}{c} V^g \epsi^{ef} \label{varV}
\en
To obtain these variations, use the
definition (\ref{liederivative}), iv) of (\ref{lieprop}),
the Bianchi identity (\ref{Rabbianchi}) and
the horizontality conditions (\ref{horizontality}).

Now we have the lemma:
\eq
\de {\cal L} = [(\de R^{ab}) V^c V^d +
          R^{ab} (\de V^c) V^d+R^{ab} V^c (\de V^d)]\epsi_{abcd}
\label{varL}
\en
\noi {\sl Proof:} use v) of (\ref{lieprop})
and the first of (\ref{epsiom}).
\sk
Inserting the variations (\ref{varR}) and (\ref{varV}) inside
(\ref{varL}) we find, after ordering the terms as
$\epsi RVV$ with (\ref{epsiom}) and (\ref{epsiR}):
\eq
\de {\cal L}=2(\Cthree{ef}{gh}{ab} \epsi_{abcd}-
\Cthree{gh}{c}{s} \epsi_{efsd}-\Cthree{rs}{d}{p} \Lthree{q}{rs}{gh}{c}
\epsi_{efqp}) \epsi^{gh} R^{ef} V^c V^d
\en
\noi Using the explicit form of the $\Lambda$ and ${\bf C}$
tensors into
(\ref{lthree})-(\ref{ctwobis}) and imposing $\de {\cal L}=0$,
we find a set of equations for the $\epsi_{abcd}$ tensor.
These can in fact be solved and yield:
\eq
\begin{array}{llll}
\epsi_{1234}=1,&\epsi_{1243}=-q,
&\epsi_{1324}=-1,&\epsi_{1342}=q,
\\
\epsi_{1423}=q^{3\over 2},&\epsi_{1432}=
-q^{3\over 2},&\epsi_{2134}=-1,&\epsi_{2143}=q,
\\
\epsi_{3124}=1,&\epsi_{3142}=-q,
&\epsi_{4123}=-q^{3\over 2},
&\epsi_{4132}=q^{3\over 2},
\\
\epsi_{2314}=q^{1\over 2},&\epsi_{2341}=
-q^{5 \over 2},&\epsi_{2413}=-q^2,&\epsi_{2431}=q^3,
\\
\epsi_{3214}=-q^{1\over 2},&\epsi_{3241}=
q^{5\over 2},&\epsi_{4213}=q^2,&\epsi_{4231}=-q^3,
\\
\epsi_{3412}=q^2,&\epsi_{3421}=-q^3,
&\epsi_{4312}=-q^2,&\epsi_{4321}=q^3 \label{qepsi}
\end{array}
\en
\sk
Consider next the variation of ${\cal L}$ under $q$-diffeomorphisms,
i.e.:
\eq
\de {\cal L}=\ell_{\epsi^g t_g} {\cal L}=
(\ell_{t_g} {\cal L}) \epsi^g - (i_{t_g} {\cal L}) d\epsi^g=
\nonumber
\en
\eq
=d[i_{t_g}(R^{ab}V^c V^d \epsi_{abcd})\epsi^g]+
i_{t_g} [d(R^{ab}V^cV^d \epsi_{abcd})]\epsi^g \label{varLdiff}
\en
\noi Then the variation $\de {\cal L}$ is a total derivative
if
\eq
d(R^{ab} V^c V^d \epsi_{abcd})=0 \label{dLiszero}
\en
\noi After using the expression for $dR^{ab}$ given by the Bianchi
identity (\ref{Rabbianchi})
 and the torsion definition (\ref{torsion}) to find $dV^a$ (note
that $R^{ab} R^c V^d \epsi_{abcd}=0$ because of horizontality
of $R^{ab},R^c$ and the vanishing of the product of five vielbeins),
eq. (\ref{dLiszero})
yields a set of conditions on $\epsi_{abcd}$. These conditions in
fact {\sl coincide} with those found to ensure local Lorentz invariance
of the $q$-lagrangian. This is not a miracle: indeed we could have
computed the Lorentz variation of ${\cal L}$ in the same way as in
(\ref{varLdiff}); the total derivative term would have vanished
because $i_{t_{gh}} (R^{ab} V^c V^d )=0$ (horizontality of $R^{ab}$),
and we would have found again eq. (\ref{dLiszero}) as a condition
for $\de {\cal L}=0$.
\sk
Thus the lagrangian (\ref{lagrangian}) with
the $\epsi_{abcd}$ tensor as given in
(\ref{qepsi}) is invariant (up to a total derivative) also
under $q$-diffeomorphisms.
\sk
We discuss now the algebra of $q$-Lie derivatives. We have
the theorem, analogous to the classical one:
\eq
\lie{i} \lie{j}-\L{kl}{ij} \lie{k} \lie{l} =
\ell_{(\C{ij}{n}-\RB{n}{ij}) t_n}  \label{liederalgebra}
\en
\noi with
\eq
R^i \equiv R^i_{~jk} \om^j \we \om^k
\en
\eq
\RB{i}{jk} \equiv R^{i}_{jk} - \L{rs}{jk} R^{i}_{rs}
\en
\noi As for the structure constants, we have
$R^i_{jk}={1\over 2} \RB{i}{jk}$
when $\Lambda^2=1$.  The proof of the
composition law (\ref{liederalgebra})
is computational: one applies its left-hand side
to $\om^k$, and uses the properties of the $q$-Lie
derivative. {\sl Hint 1}: rewrite the Lie derivative of $\om^k$ as:
\eq
\lie{j} \om^k=(\C{rj}{k}-\RB{k}{rj}) \om^r
\en

{\sl Hint 2:} use the following expression for $\L{ij}{kl}$ (no sums
 on repeated indices) :
\eq
\L{ij}{kl}= [kl] \de^i_l \de^j_k
\en
\noi and the identities
\eq
[kl]= {1 \over {[lk]}}
\en
\eq
\C{rk}{s} [lk][lr]=\C{rk}{s} [ls]
\en
\noi due to $\Lambda^2=1$ and the bicovariance condition (\ref{bic4}).
\sk
{}From the $q$-algebra (\ref{liederalgebra}) it is not
difficult to find the following composition law for
$q$-variations:
\eq
\ell_{\epsi_2^i t_i} \ell_{\epsi_1^j t_j} - (1\leftrightarrow 2)=
\ell_{(\C{ij}{n}-\RB{n}{ij})\epsi_2^i \epsi_1^j t_n}
\en
The order of the factors is important in the composite
parameter $(\C{ij}{n}-\RB{n}{ij})\epsi_2^i \epsi_1^j$. Note
also that
\eq
(\C{ij}{n}-\RB{n}{ij})\epsi_2^i \epsi_1^j=
{1 \over 2}(\C{ij}{n}-\RB{n}{ij})(\epsi_2^i \epsi_1^j-\epsi_1^i
\epsi_2^j)
\en
\noi if we postulate the commutation rule
\eq
\epsi_1^i \epsi_2^j=\L{ij}{kl} \epsi_2^k \epsi_1^l
\en
\noi Indeed $(\C{ij}{n}-\RB{n}{ij}) \L{ij}{kl}=
-(\C{kl}{n}-\RB{n}{kl})$ (due to $\Lambda^2=1$). Then
the composite parameter is explicitly $(1 \leftrightarrow 2)$
antisymmetric.
\sk
Let us derive the equations of motion corresponding
to the $q$-lagrangian (\ref{lagrangian}). We assume
the same variational rule as with the Lie derivative.
The $q$-Einstein equations are obtained by varying the
vielbein in ${\cal L}$:
\eq
\de {\cal L} = R^{ab} (\de V^c V^d + V^c \de V^d)
\epsi_{abcd}=0
\en
\noi  Postulating that $\de \om^i$ has the same commutations
as $\om^i$, and noticing that
\eq
\epsi_{abef}=-\L{cd}{ef} \epsi_{abcd}
\en
\noi (use the explicit entries in (\ref{lone}) and (\ref{qepsi}),
 or notice that since $\epsi_{abcd}$ multiplies $V^c V^d$ in
(\ref{lagrangian}), it must be $\Lambda$-antisymmetric in the
indices c,d),
one arrives at:
\eq
R^{ab} V^e \epsi_{abef}=0 \label{fieldeq1}
\en
\noi The $q$-Einstein equations are found as in the classical
case: expand (\ref{fieldeq1}) along three vielbeins:
\eq
\R{ab}{cd} V^c V^d V^e \epsi_{abef}=0,
\en
\noi multiply by another vielbein $V^g$ and use:
\eq
\epsi^{cdeg} V^1 V^2 V^3 V^4 \equiv V^c V^d V^e V^g
\en
\noi (N.B.: the entries of $\epsi^{abcd}$ are
different from those of $\epsi_{abcd}$) so that finally we have:
\eq
\R{ab}{cd} \epsi^{cdeg} \epsi_{abef}=0
\en
The contraction of the two alternating tensors yields a
$q$-weighted product of Kronecker deltas. We leave to the reader
to find the final form of the $q$-Einstein equations.
Expanding (\ref{fieldeq1}) on $\om V V$, $V \om V$ and
$\om \om V$ yields instead the horizontality condition
on $R^{ab}$.
\sk
The torsion equation is obtained by varying (\ref{lagrangian})
in the spin connection $\om^{ab}$. The final result is again an equation
that formally looks identical to the classical one:
\eq
R^c V^d \epsi_{abcd}=0
\en
As for $q=1$, this equation implies that the torsion vanishes
as a two-form: $R^c=0$ (hence also horizontality
of $R^c$).
\sk
{\sl Note 1:} The $q$-volume 4-form $V^1 V^2 V^3 V^4=
f(q) V^a V^b V^c V^d \epsi_{abcd}$ (with $[f(q)]^{-1}=
\epsi^{abcd} \epsi_{abcd}=
2q^{-{3\over 2}} (2q^{-{3\over 2}} + q^{-1} + 2q^{-{1\over 2}} + 2+
2q^{1\over 2} + q +2q^{3\over 2})$)
is invariant under Lorentz rotations and, up to a total derivative,
under $q$-diffeomorphisms. The proof is similar to the one
used for the lagrangian. This means that the $q$-symmetries
allow a cosmological constant term $V^a V^b V^c V^d \epsi_{abcd}$.
\sk
{\sl Note 2:} It would be worthwhile to give a recipe for
extracting numbers out of a $q$-theory of the kind
discussed in this Letter. A possible way of doing this
would be to find a consistent definition of path-integral on
$q$-commuting fields, leading to $\Cb$-number amplitudes.
On this question, see for example \cite{qintegral}.

\vskip 1cm
{\bf Acknowledgements}
\sk
It is a pleasure to thank the Centro Brasileiro de Pesquisas
F\'isicas of Rio de Janeiro, where this work was completed,
for its warm hospitality.

\vfill\eject

\vfill\eject
\end{document}